\documentstyle[aps,prl,epsfig]{revtex} 
\newcommand{\up}{\uparrow}
\newcommand{\dn}{\downarrow}
\begin{document}              
\twocolumn
[\hsize\textwidth\columnwidth\hsize\csname@twocolumnfalse\endcsname

\title{Pairing Correlations on $t$-$U$-$J$ Ladders }
\author{ S.\ Daul $^{1}$, D.J.~Scalapino $^{2}$ and Steven R.~White $^3$ }
\address{ 
$^1$ Institute for Theoretical Physics, University of California, \\
                  Santa-Barbara CA 93106. \\
$^2$ Physics Department, University of California, \\
                  Santa-Barbara CA 93106. \\
$^3$ Department of Physics and Astronomy, University of California, \\
	Irvine CA 92697.
}
\maketitle

\begin{abstract}              
Pairing correlations on generalized $t$-$U$-$J$ two-leg ladders are 
reported. 
We find that the pairing correlations on the usual $t$-$U$ Hubbard ladder
are significantly enhanced by the addition of a nearest-neighbor exchange
interaction $J$. 
Likewise, these correlations are also enhanced for the $t$-$J$ model when
the onsite Coulomb interaction is reduced from infinity.
Moreover, the pairing correlations are larger on a $t$-$U$-$J$ ladder than
on a $t$-$J_{\mbox{\footnotesize eff}}$ ladder in which 
$J_{\mbox{\footnotesize eff}}$ has been adjusted so that the
two models have the same spin gap at half-filling. This enhancement of the
pairing correlations is associated with an increase in the pair-binding
energy and the pair mobility in the $t$-$U$-$J$ model and point to the
importance of the charge transfer nature of the cuprate systems.
\vspace*{3mm}
\end{abstract}
]

Various {\it ab initio} quantum chemistry calculations as well as model 
Hamiltonian 
studies have been used to determine the electronic properties of Cu-oxide
clusters. \cite{Martin,McMahan,Hybertsen,Eskes,Andersen,Maekawa}
In particular, these calculations have provided parameters for simpler, 
effective one-band Hubbard
and $t$-$J$ models which have then been used to study many-body correlations in 
larger systems.
However, both the one-band Hubbard and the $t$-$J$ models differ in an
essential manner from the high $T_c$ cuprates which are known to be charge
transfer insulators \cite{Zaanen} in their undoped state. Thus, the
one-band Hubbard model at half-filling is characterized by a Mott-Hubbard
gap which is set by $U$ and in the $t$-$J$ model, $U$ is taken to infinity
with the constraint of no double occupancy. Therefore, while Coulomb
fluctuations associated with double occupancy of a site are controlled by
$U$ in the Hubbard model, $U$ also determines the strength of the exchange
coupling. In the Hubbard model as $U$ increases beyond the bandwidth, $J$
decreases as $4t^2/U$. Although $J$ is an independent parameter in the $t$-$J$
model, $U$ is infinite in this model, suppressing charge fluctuations. 

While we believe that the basic pairing mechanism arises from the exchange
correlations, the charge transfer nature of the cuprates can play an
essential role in the doped systems where it allows for a more flexible
arrangement between $J$ and $U$ than reflected in either the one-band
Hubbard or $t$-$J$ models. To explore this, we have carried out 
density-matrix renormalization group \cite{White92} (DMRG) calculations of the
pairing correlations on two-leg $t$-$U$-$J$ ladders.  Ladders are known to
provide
model systems  which exhibit various phenomena 
similar to those of the cuprates. \cite{Dag}
In particular, when doped away from half-filling they are known to have 
power-law pairing correlations which have opposite, $d_{x^2-y^2}$-like,  
signs between the rung-rung and rung-leg correlations. 
These correlations have previously been investigated for both Hubbard 
\cite{Noack,Scalapino} and $t-J$ models. \cite{Dag2,Rice}
Here we will study
a generalized $t$-$U$-$J$ model which includes both an onsite 
Coulomb repulsion $U$ and a nearest neighbor exchange $J$. 
While both Hubbard and $t$-$J$ ladders show pairing correlations when doped, we
find that these correlations can be significantly enhanced in a model with 
both $U$ and $J$. 
We argue that, in fact, a $t$-$U$-$J$ 
model is appropriate for a charge-transfer 
material. \cite{Zaanen}

The basic one-band Hubbard model is characterized by a one-electron
nearest-neighbor hopping $t$ and an onsite Coulomb interaction $U$.
\begin{equation}
   H  = \sum_{\langle ij \rangle,\sigma} -t \left( 
      c^\dag_{i\sigma} c_{j\sigma} +  c^\dag_{j\sigma} c_{i\sigma} \right)
  + U \sum_{i} n_{i\up} n_{i\dn}.
\label{eq:Hubbard} 
\end{equation}
Here $c_{i\sigma}^\dag$ creates an electron with spin $\sigma$ on site $i$ and
$\langle ij \rangle$ sums over nearest neighbor sites.
As is well known, when $U/t$ is large, a strong coupling expansion
\cite{Hirsch} of Eq. 
(\ref{eq:Hubbard}) leads to the $t-J$ Hamiltonian
\begin{eqnarray}
   H  &=& \sum_{\langle ij \rangle,\sigma} -t \left( 
      c^\dag_{i\sigma} c_{j\sigma} +  c^\dag_{j\sigma} c_{i\sigma} \right) +
   J \sum_{\langle ij \rangle} \left( {\bf S}_i {\bf S}_j 
	- \frac{n_in_j}{4} \right) \nonumber \\
&& - \frac{J}{4} \sum_{i,\delta \neq \delta',\sigma} \left( 
    c^\dag_{i+\delta,\sigma} c^\dag_{i,-\sigma} c_{i,-\sigma} 
               c_{i+\delta',\sigma}
 -   c^\dag_{i+\delta,-\sigma} c^\dag_{i,\sigma} c_{i,-\sigma} 
       c_{i+\delta',\sigma}
\right)
\label{eq:tJ} 
\end{eqnarray}
with $J=4t^2/U$ and $\delta,\delta'$ are vectors separating nearest neighbor
sites.
Here there is an important restriction that no site can have two fermions. 
Typically in Eq.. (\ref{eq:tJ}), $t$ and $J$ are treated as independent
parameters and for doping near half-filling
the latter three-site term is dropped.
Now, while these effective models both describe certain aspects of the cuprates
system, they lack the flexibility to describe an important feature that arises
from the charge-transfer nature of these materials. 
Specifically, in the insulating state the one-band Hubbard model at large $U$
has a Mott-Hubbard gap set by $U$ rather than a charge-transfer gap set by
the difference in the oxygen and copper sites energies.
Furthermore, when $U$ is large, $J\sim 4t^2/U$ decreases as $U$ increases rather
than saturating at a value set by the charge-transfer gap.
That is, in strong coupling, the three-band Hubbard model gives 
\cite{3band_strong_coupingA,3band_strong_coupingB}
\begin{equation}
      J = 4 \left( \frac{t^2_{\mbox{\footnotesize pd}}}
       {\Delta_{\mbox{\footnotesize pd}}} \right)^2 
\left[ \frac{1}{U}  + \frac{1}{\Delta_{\mbox{\footnotesize pd}}} \right]
\label{eq:J}
\end{equation}
with $t_{\mbox{\footnotesize pd}}$ the Cu ($d_{x^2-y^2}$) - O($p\sigma$)
hopping, $\Delta_{\mbox{\footnotesize pd}} $ the Cu-O site energy difference
and $U$ the Cu Coulomb energy.
There are in fact further contributions to Eq. (\ref{eq:J}) from O-O hopping
terms, as well as modifications due to O and Cu-O Coulomb interactions.
However, the basic point is that when $U$ is large compared to the effective
Cu-Cu hopping 
$t^2_{\mbox{\footnotesize pd}} / \Delta_{\mbox{\footnotesize pd}} $,
the exchange remains finite rather than going to zero.
Likewise, in the $t-J$ model, while $J/t$ can be set to a physical value,
one has in effect an infinite onsite Coulomb repulsion arising from the 
restriction of no double occupancy.
The suppression of double occupancy reduces the mobility of the pairs,  
\cite{Jeckelmann} missing the physics associated with the partial occupation 
of the O sites surrounding a Cu.

To address these limitations, we will study a $t-U-J$ model in which there is a
finite Coulomb interaction and an effective exchange term $J$. 
In the limit in which $J=0$, this is just the one-band Hubbard model while in 
the limit $U/t \gg 1$, this goes over to the $t-J$ model \cite{Ref}
The DMRG calculations reported here have been carried out on open ended 
ladders (up to $2 \times 48$ sites)
keeping up to 800 states, so that the maximum weight of the discarded density 
matrix eigenvalues is $10^{-6}$. 
We first examine the rung-rung pair-field correlation function 
\begin{equation}
     D(\ell) = \langle  \Delta_{i+\ell}  \Delta^\dag_{i}   \rangle
\end{equation}
for a doped (8 holes) $2 \times 32$ ladder.
The operator
\begin{equation}
     \Delta^\dag_{i}  = c^\dag_{i1,\up} c^\dag_{i2,\dn} -
                          c^\dag_{i1,\dn} c^\dag_{i2,\up}
\end{equation}
creates a singlet pair on the $i^{\mbox{{\footnotesize th}}}$ rung and 
$\Delta_{i+\ell}$ 
destroys it on the $ (i + \ell) ^{\mbox{{\footnotesize th}}}$ rung.
A similar calculation in which a singlet pair is created on the  
$i^{\mbox{{\footnotesize th}}}$ rung and a singlet pair is destroyed on one 
of the legs at $i+\ell$ has an 
opposite sign indicating the $d_{x^2-y^2}$-like structure of the pairing.
Because of the finite length of the ladder, we have kept $\ell \leq 12$,
with the measurements made in the central portion of the ladder,
in the plots of $D(\ell)$.
In this region the effects of the open ends are negligible.

\begin{figure}[htb]
 \begin{center}
  \epsfig{file=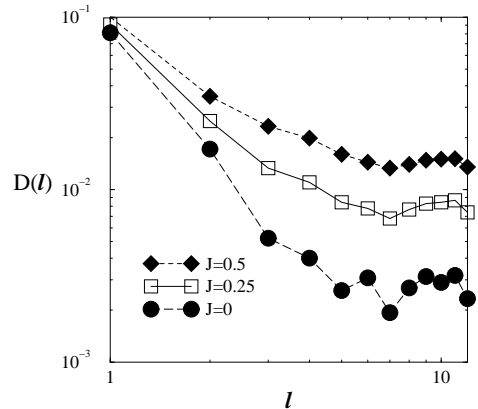,width=6cm}
 \end{center}
 \caption{The rung-rung singlet pairing correlation function
$D(\ell)$ versus $\ell$ on a doped $2 \times 32$ ladder with 
$\langle n \rangle = 0.875$ for $U=6$ and various values of $J$. }
 \label{fig:dyy_J}
\end{figure}

In Fig. \ref{fig:dyy_J} we show the effect of adding an additional 
exchange term $J$ to a Hubbard model with $U=6$.
Here and in the following we will measure energy in units of $t$. 
As seen, the addition of $J$ clearly enhances the pairing. 
In all of the plots it is important to recognize that the pair has an internal
structure so that $\Delta^\dag_{i} $ and $\Delta_{i+\ell} $ have only a 
partial overlap to the state in which a pair is added at the 
$i^{\mbox{{\footnotesize th}}}$ rung or removed from the $i+\ell$ rung, 
and the basic size of $D(\ell)$ is reduced by the square of this overlap.
As seen in Fig.~1, adding an additional exchange strongly enhances the
pair-field correlations.

\begin{figure}[htb]
 \begin{center}
  \epsfig{file=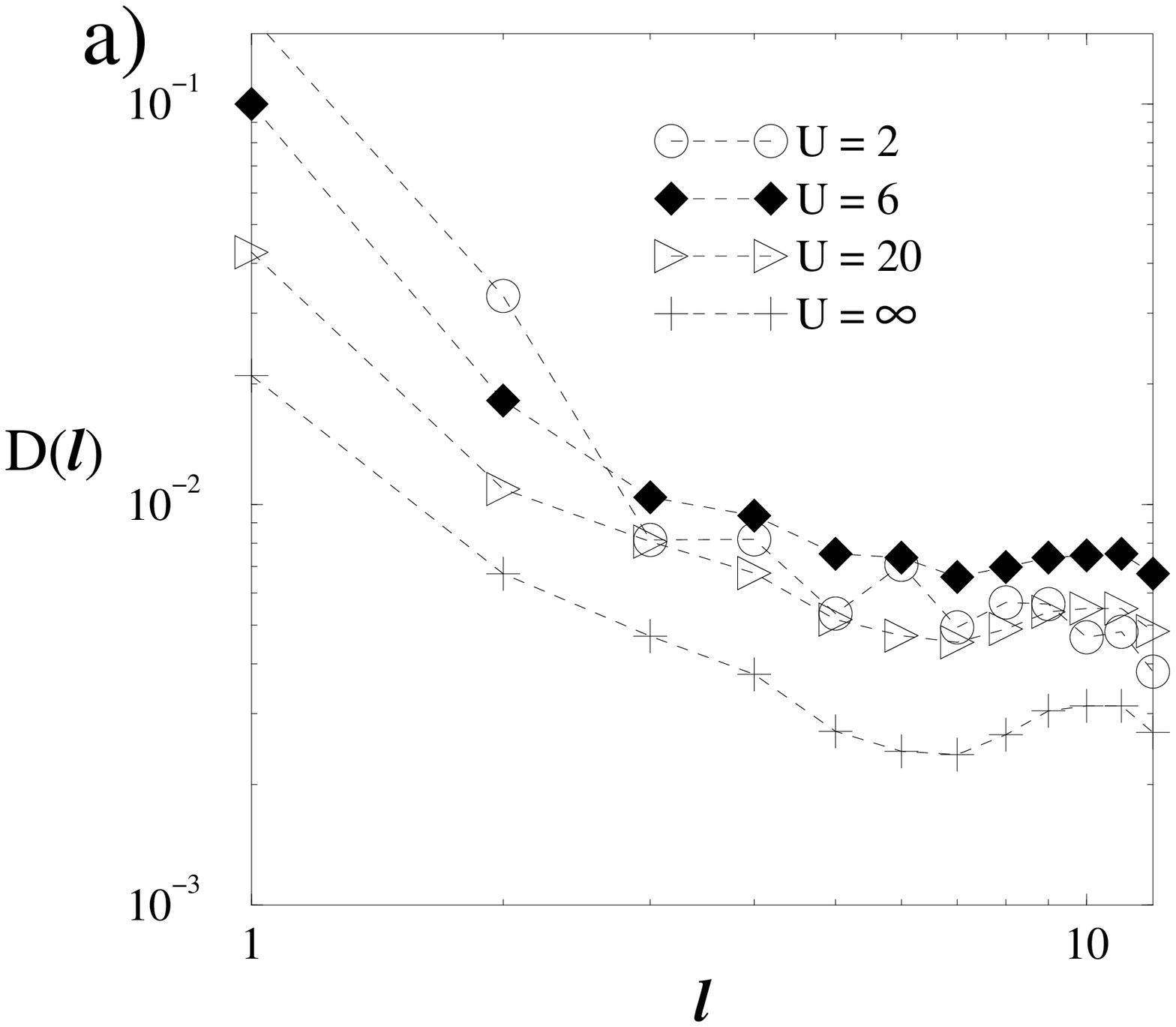,width=6cm}
  \epsfig{file=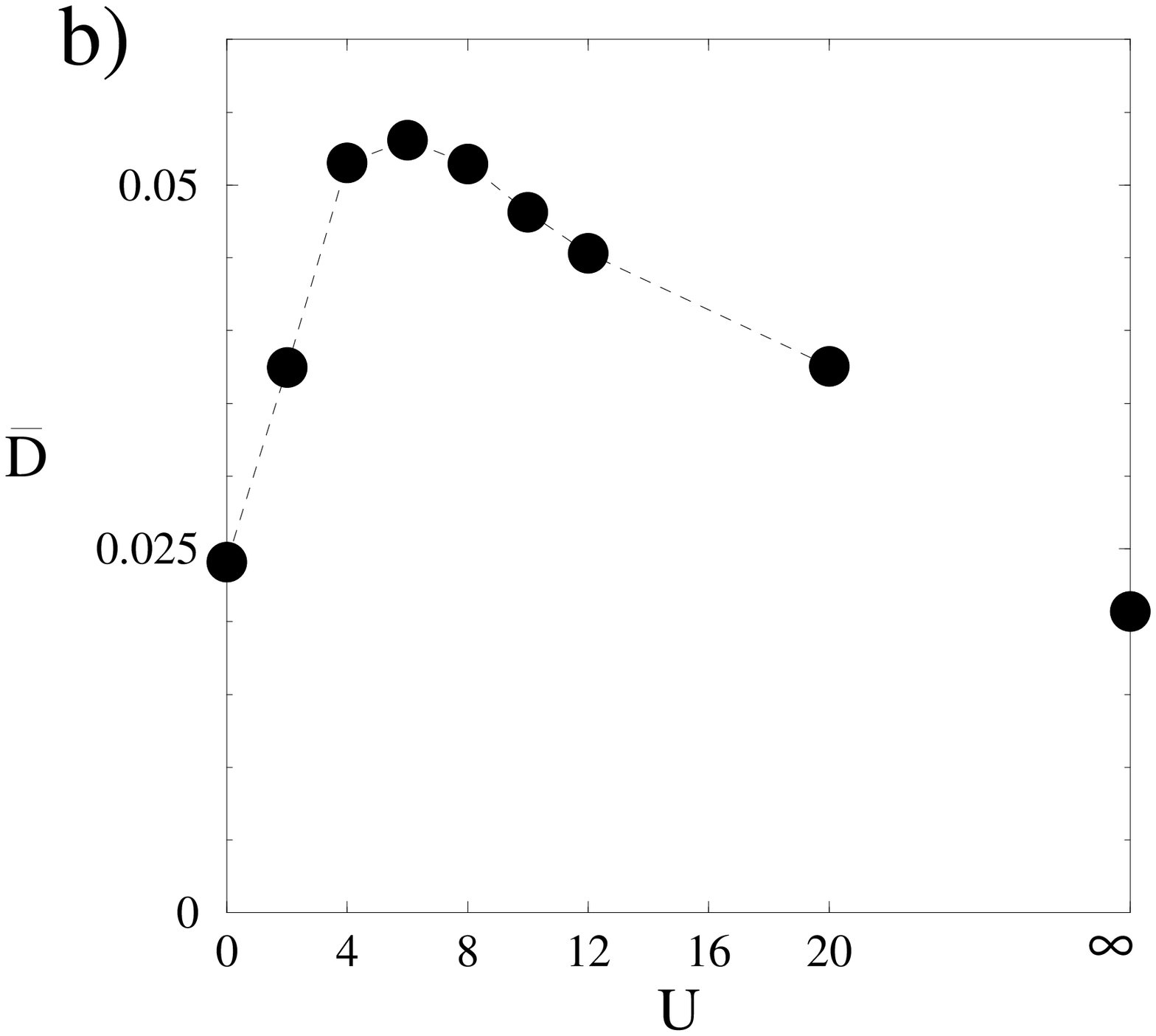,width=6cm}
  \label{fig:dbar_U}
 \end{center}
 \caption{(a) The rung-rung singlet pairing correlation function
$D(\ell)$ versus $\ell$ on a doped ladder with $\langle n \rangle = 0.875$ for
$J=0.25$ and various values of $U$.
(b) The partial singlet pairing correlation function sum $\bar{D}$ as a 
function of $U$ for a doped ladder with $\langle n \rangle = 0.875$ 
and $J=0.25$.}
  \label{fig:dyy_U}
\end{figure}

Similarly, in Fig.~\ref{fig:dyy_U}a, we examined the effect of $U$ on the 
pairing correlations of a $t$-$U$-$J$ ladder with $J=0.25$.
For $U \gg 1$, we have the usual $t-J$ result.
As $U$ initially decreases, there is again a significant enhancement of the 
pairing
correlations, but eventually as $U$ decreases below the bandwidth, the
pairing correlations are reduced. This is also shown in Fig.~\ref{fig:dyy_U}b,
where we have plotted
\begin{equation}
\bar D=\sum^{12}_{\ell=8} D(\ell)
\label{six}
\end{equation}
versus $U$ for $J=0.25$. Here $\bar D$ reaches a maximum for $U\simeq 6$.

One would, of course, expect that the pairing correlations would depend on the
total effective exchange interaction, both the explicit ``$J$'' exchange
and the additional exchange associated with a finite $U$.
Thus, in the $t$-$U$-$J$  model, as $U$ initially
increases, the effective exchange increases and then as $U$ exceeds the
bandwidth its contribution to the exchange decreases as $4t^2/U$.  However,
there is more to this than just the enhancement of the exchange interaction
which can be seen by comparing the two models.  A half-filled Hubbard
ladder with $U=6$ and $J=0.25$ has a spin gap $\Delta_s=0.22$ corresponding
to an effective exchange \cite{SpinGap}
 $J_{\mbox{\footnotesize eff}} \approx 2\Delta_s=0.44$. 
Using this value for the exchange in a $t-J$ model we have calculated
the pair-field correlation function $D(\ell)$ in Fig.~3 and  compared it with
the pair-field correlations found for the corresponding $t-U-J$ model. 
Although both of these models have the same spin gap at half-filling, it is
clear that the $t-U-J$ ladder has significantly stronger pairing correlations.
\begin{figure}[thb]
\begin{center}
  \epsfig{file=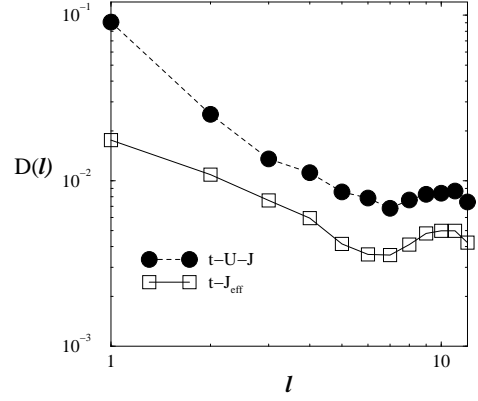,width=6cm}
 \end{center}
 \caption{ Comparison of $D(\ell)$ versus $\ell$ for a $t$-$U$-$J$ model
(solid) with $U=6$ and $J=0.25$ with a $t$-$U$ model (open) which has the
same spin gap at half-filling.}
\end{figure}

\begin{figure}[htb]
 \begin{center}
  \epsfig{file=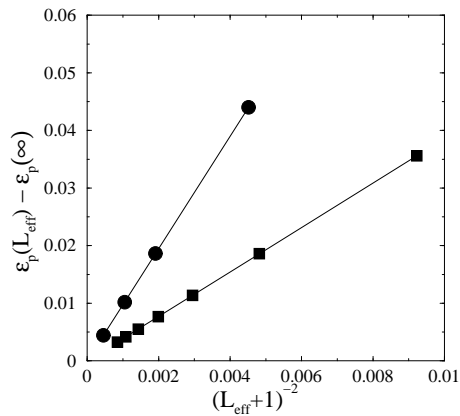,width=6cm}
 \end{center}
 \caption{Hole pair energy $\epsilon_p$ versus 
  $ \left( L_{\mbox{\footnotesize eff}} + 1  \right)^{-2}$ 
for the $t-U-J$ model with $U=6$ and $J=0.25$ (circles) and the corresponding 
$t-J$ model with $J=0.44$  (squares).  The solid lines are least mean square
fit of Eq. (\ref{eq:pairgap}). }
 \label{fig:pairgap}
\end{figure}

In order to understand the reasons for this, we have calculated the 
pair-binding energy and the pair mobility for both these models. 
The pair-binding energy is
\begin{equation}
  E_{\mbox{\footnotesize pb}} = 2 E_0(1) - E_0(2) - E_0(0) 
\label{seven}
\end{equation}
with $E_0(n)$ the ground-state energy with $n$ holes.
We find $E_{\mbox{\footnotesize pb}}$  is equal to $0.34$
for the $t$-$U$-$J$ model with $U=6$ and $J=0.25$.  For the 
$t$-$J_{\mbox{\footnotesize eff}}$
ladder with $J_{\mbox{\footnotesize eff}}=0.44$, adjusted so that the two 
models have the same
spin gap at zero doping, the pair-binding energy is $0.23$. 
We have also calculated the effective hopping $t_{\mbox{\footnotesize eff}}$
of a hole pair from the dependence of 
\begin{equation}
  \epsilon_p (L_x)  = E_0(2) - E_0(0)
\end{equation}
on the length of the ladder for ladders with $L_x$ up to  48.
In ladders with open boundary conditions,  $\epsilon_p(L_x)$ varies as
\begin{equation}
  \epsilon_p (L_x) =  \epsilon_p(\infty) + t_{\mbox{\footnotesize eff}}
  \;\frac{\pi^2}{\left(L_{\mbox{\footnotesize eff}} + 1  \right)^2}
\label{eq:pairgap}
\end{equation}
where the effective length differs from the actual ladder length $L_x$
because of end effects.
For large enough systems, the difference 
$L_{\mbox{\footnotesize eff}} - L_x = \delta L$ tends to a constant and 
is considered as a fitting parameter. \cite{Jeckelmann}
Fig. \ref{fig:pairgap} shows the results for the $t$-$U$-$J$ and
the $t-J_{\mbox{\footnotesize eff}}$ models. 
The effective hopping, given by the slope divided by $\pi^2$, is 
$t_{\mbox{\footnotesize eff}}=0.99$ for the $t$-$U$-$J$ ladder and 
$t_{\mbox{\footnotesize eff}}=0.39$ for the 
$t-J_{\mbox{\footnotesize eff}}$ ladder.

The enhancement of the effective pair hopping which occurs when $U$ is finite
can be understood as arising from virtual states involving doubly occupied 
sites. 
An example of this is illustrated in Fig.~5. Here a pair of holes on the top
rung hops to the bottom rung via a set of intermediate states. 
In this sequence, the second intermediate state, shown in the middle of 
the figure, has a doubly occupied site. 
In the $t-U-J$ model this would not be allowed, leading to a reduction
in the effective pair hopping.
This effect not only enhances the pair-field correlations on the $t-U-J$
ladder, but we believe also would act to reduce the stripe stiffness in the
2D $t-J$ problem. This would favor a $d_{x_2-y_2}$-pairing state over the 
striped state we have typically found in DMRG calculations on $n$-leg
$t-J$ ladders. \cite{White2}
\begin{figure}[ht]
 \begin{center}
  \epsfig{file=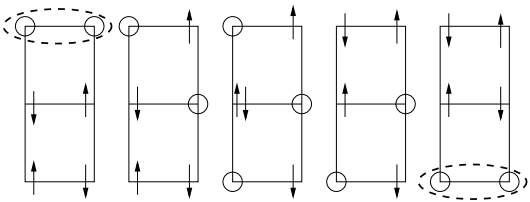,width=7cm}
 \end{center}
 \caption{Illustration of a pair transfer process involving a set of
        intermediate states. The state in the center has a doubly
        occupied site. This transfer process cannot occur in the
        $t-J$ model but it does contribute to the pair hopping in
        the $t-U-J$ model.}
 \label{fig:5}
\end{figure}

Thus we conclude that the charge transfer nature of the cuprates can
be more appropriately described using a $t-U-J$ model. Furthermore, this
model exhibits enhanced pairing correlations due to (1) an additional
exchange coupling reflecting the exchange path in which there is a virtual
double occupancy on the oxygen rather than the Cu and (2) an enhanced pair
hopping allowed by a finite value of $U$ which reflects the alternate
paths for electron transfer in the charge transfer system.

We wish to thank D.~Duffy and M.~Fisher for helpful discussions. 
SD acknowledges support from the Swiss National Science Foundation and the
ITP under NSF grant PHY94-07194. 
DJS and SRW wishes to acknowledge partial support from the US Department of
Energy under Grant No. DE-FG03-85ER45197.




\begin{references}

\bibitem{Martin} R. L. Martin, J. Chem. Phys. {\bf 98}, 8691 (1993).

\bibitem{McMahan} A. K. McMahan, J. F. Annett and R. M. Martin, Phys. Rev.
B {\bf 42}, 6268 (1990). 

\bibitem{Hybertsen} M. S. Hybertsen, E. B. Stechel, M. Schluter and
D. R. Jennison, Phys. Rev. B {\bf 41}, 11068 (1990).

\bibitem{Eskes} H. Eskes, L. H. Tjeng and G. A. Sawatzky, Phys. Rev. B 
{\bf 41}, 288 (1990).

\bibitem{Andersen} O. K. Andersen, O. Jepsen, A. I. Liechtenstein and
I. I. Mazin, Phys. Rev. B {\bf 49}, 4145 (1994).

\bibitem{Maekawa} T. Tohyama and S. Maekawa, Journ. of the Phys. 
Soc. Jap. {\bf 59}, 1760 (1990).

\bibitem{Zaanen} J. Zaanen, G. A. Sawatsky and J. W. Allen, Phys. Rev.
Lett.
{\bf 55}, 418 (1985).

\bibitem{White92} S. R. White, Phys. Rev. Lett. {\bf 69}, 2863 (1992);
Phys. Rev. B {\bf 48}, 10345 (1993).

\bibitem{Dag} E. Dagotto and T. M. Rice, Science {\bf 271}, 618 (1996).

\bibitem{Noack}  R. M. Noack, D. J. Scalapino and S. R. White, Phil.
Mag. {\bf 74}, 485 (1996). 

\bibitem{Scalapino} R. M. Noack, S. R. White and D. J. Scalapino, 
Phys. Rev. Lett. {\bf 73}, 882 (1994).

\bibitem{Dag2} E. Dagotto, J. Riera and D. J. Scalapino, Phys.
Rev. B {\bf 45}, 5744 (1992). 

\bibitem{Rice} T. M. Rice, M. Troyer and H. Tsunetsugu, J. Phys. Chem.
Solids {\bf 56}, 1663 (1995).

\bibitem{Hirsch} J. E. Hirsch, Phys. Rev. Lett. {\bf 54}, 1317 (1985).

\bibitem{3band_strong_coupingA} V. Emery, Phys. Rev. Lett. {\bf 58}, 
2794 (1987).

\bibitem{3band_strong_coupingB} H. Eskes and J. H. Jefferson,
Phys. Rev. B {\bf 48}, 9788 (1993).

\bibitem{Jeckelmann} E. Jeckelmann, D. J. Scalapino and S. R. White, 
Phys. Rev. B {\bf 58}, 9492 (1998).

\bibitem{Ref} We have
also studied the effects of the 3-site hopping term in eq.~(\ref{eq:tJ})
 as well as additional next-nearest-neighbor hopping $t^\prime$ and
next-next-nearest-neighbor hopping $t^{\prime\prime}$ terms, but
these results will be reported elsewhere.


\bibitem{SpinGap}  S. R. White, R. M. Noack and D. J. Scalapino,
Phys. Rev. Lett. {\bf 73}, 886 (1994). 

\bibitem{White2} S. R. White and D. J. Scalapino, Phys. Rev. B {\bf 60}, R753
(1999).



\end{references}
\end{document}